# Einstein was right: Proof of absence of spooky state reduction in quantum mechanics

C. S. Unnikrishnan

*Gravitation Group, Tata Institute of Fundamental Research, Homi Bhabha Road, Mumbai - 400 005, India &*

*NAPP Group, Indian Institute of Astrophysics, Bangalore - 560 034, India.*

E-mail addresses: unni@tifr.res.in

**The present standard interpretation of quantum mechanics invokes nonlocality and state reduction at space-like separated points during measurements on entangled systems. While there is no understanding of the physical mechanism of such nonlocal state reduction, the experimental verifications of quantum correlations[1] different from that predicted by local realistic theories have polarized the physicists' opinion in favour of nonlocality. I show conclusively that there is no such spooky state reduction, vindicating the strong views against nonlocality held by Einstein[2] and Popper[3]. Experimental support[4] to this proof is also discussed. The Bell's inequalities arise due to ignoring the phase information in the correlation function and not due to nonlocality[5]. This result goes against the current belief of quantum nonlocality held by the majority of physicists; yet the proof is transparent and rigorous, and therefore demands a change in the interpretation of quantum mechanics and quantum measurements. The hypothesis of wave function collapse is inconsistent with experimental observations on entangled correlated systems.**

Quantum nonlocality contained in the supposed state reduction at a distance during measurements on entangled quantum systems goes against the spirit of relativity[6] and much of rational physics[3]. Interestingly, there is no direct proof from any experiment that there is indeed a nonlocal influence causing nonlocal state reduction. In fact, there are indications that there is no superluminal physical influence that passes between the separated subsystems of the entangled state[5,7]. What the experiments show is that the observed quantum correlations are stronger than what is predicted by a general local hidden variable theory[1]. Physicists have by and large concluded that this indicates that there is nonlocality in quantum mechanics – state reduction without a real local measurement, as a result of another measurement at a space-like separated region.

In the language of the general formalism of quantum mechanics, the initial entangled state is a superposition of multiparticle product states,

$$\Psi = \frac{1}{\sqrt{N}} \sum_i |x_i\rangle_1 |x_i'\rangle_2$$

The summation is over a finite number of product states of the two particles such that the total state cannot be written as a simple product state. A measurement on even one particle reduces this state to $\Psi_{12} = |x_j\rangle_1 |x'_j\rangle_2$ with probability 1/N, where *j* is one of the possible eigenvalue labels. The pairs of eigenvalues $\{x_j, x'_j\}$ defines the quantum correlation. After the measurement, the state of the first subsystem is reduced to $\Psi_1 = |x_j\rangle$ and then the present interpretation of quantum mechanics is taken to imply that the factual quantum state of the second subsystem is instantaneously and nonlocally reduced to the state $\Psi_2 = |x'_j\rangle$, even if it is space-like separated from the first at an arbitrarily large distance. This is how nonlocality enters quantum mechanics.

Now we show that this firmly held belief is incorrect and that there is no state reduction at a distance. The standard interpretation is contradicted by the mathematical prediction of quantum mechanics itself, and also by experiment. The proof is general and simple. To prove this we start with a fundamental lemma of quantum mechanics.

A) Fundamental Lemma : An eigenstate of an observable is in general a superposition of eigenstates of another noncommuting observable. In particular, an eigenstate of position is a superposition of eigenstates of momentum.

$$\psi(x) = \frac{1}{(2\pi\hbar)^{1/2}} \int_{-\infty}^{\infty} dp \exp(ipx/\hbar) \phi(p)$$

The dispersion (spread) in momentum δp for a quantum state localized in position (an eigenstate of position) with uncertainty δx is given by δp≥ħ/δx.

B) Proposition (from standard interpretation of quantum mechanics): For an Einstein-Podolsky-Rosen type of entangled state[2], with position eigenvalues satisfying the correlation *x1+ x2=0*, the state reduction of the first particle to the position eigenstate $|x1\rangle$ with eigenvalue *x1*, implies that the second particle is instantaneously and nonlocally collapsed to the eigenstate of position $|x2\rangle$ with eigenvalue *–x1*.

Our proof is to show that (B) is contradicted by (A). Consider an EPR-like state, with entanglement and correlation in transverse position and momentum. If the first particle is measured to be in one of the eigenstates of position, it will obey the fundamental lemma. This can be observed in the experiment as a spreading of the probability for its subsequent detection, or equivalently as a dispersion in the transverse momentum obeying the uncertainty relation. The fundamental lemma then implies that *if the second particle is reduced to an eigenstate of position, then it is also a superposition of momentum states, exactly as for the first particle.* Equivalently, the behaviour of the second particle will have to be identical to that of the first particle after the measurement on the first particle. In other words, *the probability of detection of the second particle should start spreading spontaneously, immediately after the measurement on the first particle.*

This conclusion derived from proposition B and the fundamental lemma is in conflict with the predictions of quantum mechanics, and also with the requirement signal locality. Quantum mechanics predicts that the second particle, if measured, will be found in an eigenstate of position (with an uncertainty given by the angular size of the slit), *irrespective of the distance at (or delay after) which this measurement is made.* Therefore, the reduction to an eigenstate *has to wait till the real measurement is made.* This is because a position eigenstate does not remain as a position eigenstate under evolution; it disperses. Conversely, if it does not, then it cannot be in an eigenstate of position. Thus, the simple mathematical and physical criterion for state reduction is violated by the 'unmeasured' particle in the entangled pair.

*Therefore proposition B is false. There is no nonlocal state reduction.*

If the dispersion in the momentum for the second particle changed as a result of the measurement on the first particle, then signal locality can be violated statistically[8]. The specific protocol would involve making the measurements with two different slits in sequence, each slit measuring a large number of particles from the entangled ensemble, sufficient in number to identify a 'high' or 'low' in dispersion in momentum of the second particle.

Similar arguments can be made for experiments with particles entangled in energy and time variables, with entangled state $\delta(E_1 + E_2 - E_0)$ instead of the EPR state $\delta(x_1 + x_2)$. The difficulty in doing experiments using entanglement in spin variables, due to the bounded nature of dispersion in values of spin projection, is pointed out in Ref. 5.

We depict the drastic conflict between the present standard interpretation implying nonlocal state reduction, and the real mathematical prediction of quantum mechanics in Figure 1. This also suggests how to perform an experiment that will directly test whether there is nonlocal state reduction in quantum mechanics. The experiment involves measurement of the transverse position of one particle in a pair of particles entangled in position and momentum. If the particle is detected behind a narrow slit, then it is also localized within the size of the slit. Then, according to standard quantum mechanical interpretation with nonlocal state reduction, the second particle is instantaneously localized to a correlated position. Since the second particle is available for a later measurement, it can be directly checked in an ensemble of such measurements whether it is in the Fourier sum of momentum states, demanded by the fundamental Lemma. If it is, then there is a finite nonzero probability for its detection at positions far away from the correlation defined by the relation *x1+ x2=0. But the mathematical formalism of quantum mechanics definitively predicts that the relation x1+ x2=0 is obeyed even if the second particle is allowed to evolved for some arbitrary finite time before the measurement on the second particle is made.*

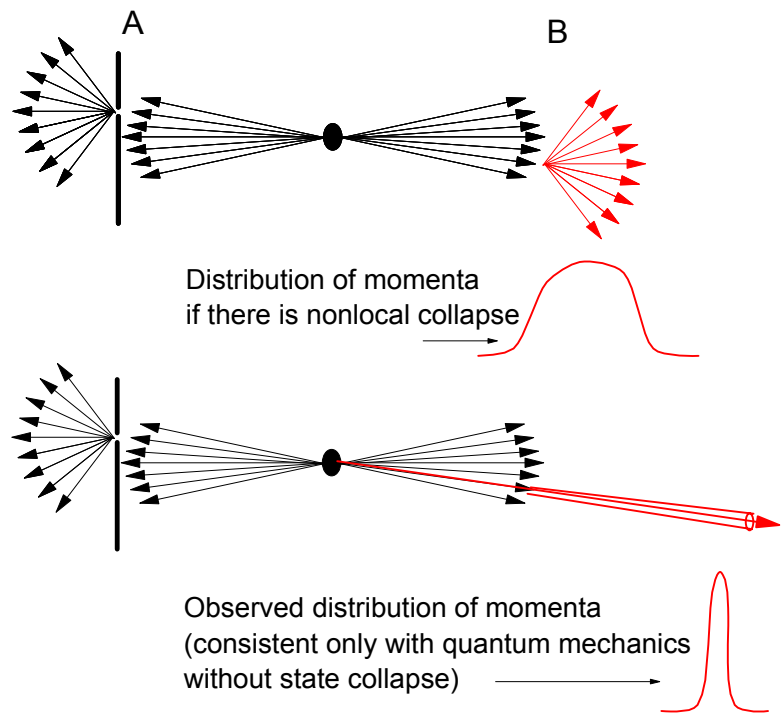

*Fig. 1: The upper panel shows that the momentum dispersion of the first particle increases after its state is reduced to an eigenstate of position, and that similar dispersion should happen to the second particle instantaneously **if** its state is nonlocally reduced to a particular eigenvalue. The lower panel shows the mathematical prediction of quantum mechanics, in which the probability of detecting the particle remains constant, within a fixed solid angle. Experiment and arguments using signal locality confirms the latter, ruling out nonlocal state reduction.*

It turns out that such an experiment was already done in a different context. In fact, an experiment that uses the same configuration was proposed by K. Popper[9] in support of his criticism of the Copenhagen interpretation. We can use the result of such an experiment for our purpose of experimentally refuting nonlocality since the experimental configuration is similar. In the experiment that realized the configuration discussed above[4] the second particle's momentum dispersion was found to be less than what a state reduction and localization of the second particle imply. In other words, the dispersion for the second particle violated the bound given by the superposition principle and the fundamental lemma. <u>So, experimental results show that there is no state reduction at a distance.</u>

Therefore, the general belief of nonlocal quantum state reduction, *derived from the projection postulate that is external to the basic mathematical machinery of quantum mechanics*, is shattered by being in conflict with the quantum mechanical prediction itself! More importantly, the hypothesis of nonlocal state reduction is rejected by experiments.

The proof is completely independent of any specific definition of state reduction since we arrived at the proof by comparing the behaviour of the two particles of the entangled system, one subjected to a real measurement and another one that was supposed to be reduced to a similar definite eigenstate even without a real measurement.

This proof clearly indicates that it is not the assumption of locality in Bell's theorem that is to be abandoned. <u>A careful examination of the proof of the Bell's inequalities reveal that the Bell correlation function does not take into account of quantum information that is encoded in the phase of the individual local amplitudes[5]. By multiplying the eigenvalues and then averaging to get a correlation function, the crucial phase information that is essential for getting the correct correlations is left out</u>. This is not physically correct as examples from optics (like the Hanbury Brown-Twiss intensity correlations) show. Imagine calculating an interference pattern by combining only the relevant eigenvalues without phases. An inequality is sure to emerge that is violated by experiments. The situation regarding correlations is similar. If one starts with a local description that incorporates phase, then the correct quantum correlations can be obtained[5].

This proof decisively settles the debate on nonlocality in quantum measurements on entangled systems in favour of the views expressed by Einstein and Popper. It shows that the present interpretation of quantum mechanical state reduction applied to entangled systems that are separated into space-like regions should be modified. A correct interpretation would be that the state vector represents what can be known of the physical system *if* a measurement is made, and not what the physical system is. Just as the *a priori* probabilities could differ from *a posteriori* probabilities after measurements on correlated classical statistical systems, *a posteriori* probability amplitudes are in general different from *a priori* probability amplitudes. But measurement does not lead to any nonlocal change in the quantum system. Also, this proof is in perfect harmony with the decoherence view of measurements since decoherence is a local process due to interaction with an environment. The proof that there is nonlocal state reduction removes the discord between the spirit of special relativity and quantum mechanics.

Acknowledgments: I thank E. C. G. Sudarshan and R. Cowsik for several discussions and encouragement. I thank F. Lalöe and R. Simon for critical comments that helped to sharpen the proof.